\documentclass[a4paper,english,twocolumn,amsmath,amssymb,superscriptaddress]{revtex4}
\usepackage[T1]{fontenc}
\usepackage[latin1]{inputenc}
\usepackage{babel}
\usepackage{array}
\usepackage{graphicx}

\makeatletter

\begin{document}

\title{Homogeneous nucleation under shear in a  two-dimensional Ising model: cluster growth, coalescence and breakup}

\author{Rosalind J. Allen}
\affiliation{SUPA, School of Physics, The University of Edinburgh, 
Mayfield Road, Edinburgh EH9 3JZ, UK.}
\author{Chantal Valeriani}
\affiliation{FOM Institute for Atomic and Molecular Physics,
Kruislaan 407, 1098 SJ Amsterdam, The Netherlands.}
\affiliation{SUPA, School of Physics, The University of Edinburgh, 
Mayfield Road, Edinburgh EH9 3JZ, UK.}
\author{Sorin T{\u{a}}nase-Nicola}
\affiliation{FOM Institute for Atomic and Molecular Physics,
Kruislaan 407, 1098 SJ Amsterdam, The Netherlands.}
\affiliation{University of Michigan, Physics Dept.,  450 Church St., Ann Arbor, MI 48109-1040, USA}
\author{Pieter Rein ten Wolde}
\affiliation{FOM Institute for Atomic and Molecular Physics,
Kruislaan 407, 1098 SJ Amsterdam, The Netherlands.}
\author{Daan Frenkel}
\affiliation{FOM Institute for Atomic and Molecular Physics,
Kruislaan 407, 1098 SJ Amsterdam, The Netherlands.}
\affiliation{Department of Chemistry, University of Cambridge, Lensfield Road, Cambridge CB2 1EW, UK.}

\date{\today}

\begin{abstract}
We compute rates and pathways for nucleation in a sheared two dimensional Ising model with Metropolis spin flip dynamics, using Forward Flux Sampling (FFS). We find a peak in the nucleation rate at intermediate shear rate. We analyse the origin of this peak using modified shear algorithms and committor analysis. We find that the peak arises from an interplay between three shear-mediated effects: shear-enhanced cluster growth, cluster coalescence and cluster breakup. Our results show that  complex nucleation behaviour can be found even in a simple driven model system. This work also demonstrates the use of FFS for simulating rare events, including nucleation, in nonequilibrium systems.
\end{abstract}
\maketitle
\section{Introduction}

The nucleation of a stable phase from a metastable one is a ubiquitous and important phenomenon. Most progress in understanding the physics of nucleation has been made for  ``quasi-equilibrium'' systems, in which the system dynamics obeys detailed balance and the transition is from a metastable to a thermodynamically stable state. However, many important nucleation processes both in nature and in industry happen in driven systems, such as those under shear, whose dynamics do not obey detailed balance. Despite its importance, nucleation in driven systems remains poorly understood. In this paper, we compute rates and transition paths for a driven nucleation process: nucleation under shear in a two dimensional Ising model. We use the recently developed Forward Flux Sampling rare event simulation method \cite{FFS,FFS2,FFS3}. We find that the nucleation rate shows a striking nonmonotonic dependence on the shear rate, and that this is due to an interplay between three shear-mediated effects: shear-enhanced cluster growth, cluster coalescence and cluster breakup. 

Nucleation under shear  remains poorly understood \cite{vermant,onuki}.  It is expected that high enough shear rates will impede nucleation. Some studies of crystal nucleation \cite{butler,blaak1,blaak2}  find that nucleation rates decrease monotonically with shear rate; others suggest that  crystallisation may be enhanced at low shear rates \cite{cerda,gray,ackerson1,haw}. A recent experimental study found a minimum in the crystal nucleation rate as a function of shear rate for charged colloids \cite{holmqvist}. Crystallisation from sheared glassy states is even more complicated, both experimentally and numerically \cite{haw,mokshin}. For binary mixtures \cite{chan} and isotropic-to-lamellar transitions \cite{cates}, shear is predicted to increase the critical temperature. Physical mechanisms for the effect of shear on nucleation may include hydrodynamic effects, cluster coalescence, cluster breakup, layering due to the shear flow, and suppression of polydispersity. In this work, we study an idealised model in which many of these effects are not included (perhaps most significantly, transport processes are not modelled). Our motivation is to provide a fundamental basis on which to build an understanding of more complex systems. Our results may however be relevant to driven solid materials \cite{martin,bellon}.
  
 The Ising model provides a paradigm for many phenomena in statistical physics, including nucleation. Nucleation in this model, in the absence of external driving, has been extensively studied \cite{binder1,schneidman,shn2,coniglio,abraham,achar,3dim_old,barkema,wonczak,sear}. Ising models have proved a valuable tool for testing the Classical Nucleation Theory (CNT)\cite{Becker,nucl_old1, nucl_old2}, in which nucleation is coarse-grained to one dimension, the largest cluster size coordinate, and modelled as a transition over a free energy barrier that arises due to competition between the favourable chemical potential of the growing cluster and its unfavourable interfacial free energy. An attempt has been made to extend the CNT to sheared systems \cite{reguera}. In the absence of shear, transition path analysis has shown the importance of order parameters other than the largest cluster size in the nucleation mechanism, in both two and three dimensions \cite{pan_jpcb2004,peters1}. This large body of information on nucleation in the absence of driving makes the Ising model an attractive test system for nonequilibrium nucleation problems. Metastability and nucleation of nonequilibrium steady states generated by coupling to two different heat baths has been studied in a two-dimensional Ising model \cite{Hurtado2004,Hurtado2006}. Although, to our knowledge, nucleation under shear has not been investigated for the Ising model,  the maximum likelihood path has recently been computed for nucleation under shear in a finite system defined by a dynamical equation for the nucleation order parameter, in the absence of applied field \cite{Heymann}. In this paper, we study a sheared two-dimensional Ising model. We find that even for this highly simplified system, nucleation under shear is a complex process that depends on multiple physical mechanisms.

The Forward Flux Sampling (FFS) method used in this work allows the computation of rate constants, transition paths and stationary probability distributions for rare events in equilibrium or nonequilibrium systems \cite{FFS,FFS2,barrier}. Rare events, such as nucleation, are notoriously difficult to simulate, because the waiting time between events is typically much longer than the timescale of the event itself, meaning that few, if any, events are observed in a typical simulation run. Rare event simulation methods developed for equilibrium systems include Bennett-Chandler methods \cite{chandler}, transition path sampling \cite{dellago2}, (partial path) transition interface sampling \cite{vanerp1},  milestoning \cite{faradjian04} and string methods \cite{Ren02}. Both TPS and the string method have been applied to Ising nucleation \cite{pan_jpcb2004,peters1,Venturoli}. However, these methods require knowledge of the steady state phase space density, making them unsuitable for nonequilibrium problems. FFS does not require knowledge of the phase space density. The method uses a series of interfaces in phase space between the initial and final states, defined by an order parameter which need not be the reaction coordinate. In earlier work, we and others have shown that FFS correctly reproduces the nucleation behaviour of a two dimensional Ising model in the absence of shear \cite{barrier,sear}.

In section \ref{sec:sim}, we give details of our simulation model and the FFS method applied to this system. In section \ref{sec:rate}, we present results for the nucleation rate as a function of the shear rate. We then analyse the physical mechanism behind the suppression of nucleation at high shear rates in section \ref{sec:supp}. In section \ref{sec:enh}, we discuss the roles of shear-enhanced cluster growth and coalescence in the enhancement of nucleation at low shear rates. We test our ideas with an analysis of cluster growth in section \ref{sec:growth}, with a comparison to a modified shear algorithm in section \ref{sec:mod}, and with an analysis of the transition state ensemble in section \ref{sec:antse}. Finally, we present our conclusions in section \ref{sec:dis}.

\section{Simulation Details}\label{sec:sim}
\subsection*{The sheared two-dimensional Ising model}
Our system consists of a two-dimensional  $L\times L$ square lattice of up-down spins with nearest-neighbour spin-spin interactions, coupling to an external magnetic field, and periodic boundary conditions in the $x$ and $y$ directions. The Hamiltonian for the spin-spin and spin-field interactions is
\begin{equation}
H = - J \sum_{ij}^{\hspace{0.2cm}  '} \sigma_{i} \sigma_{j} - h \sum_{i} \sigma_{i},
\label{eq:ham}
\end{equation}
where $\sigma_{i} = \pm 1$ is the state of spin $i$, $J$ is the coupling constant between neighbouring spins and $h$ the external magnetic field - both in units of $k_BT$.  The prime indicates that the sum is restricted to nearest-neighbour interactions. Our simulations use  Metropolis spin-flip dynamics. In each Monte Carlo cycle, we make $L\times L$ attempts to flip a spin. In each attempt, we choose a spin at random, attempt to flip it, and accept or reject the flip according to the Metropolis rule. An alternative choice of dynamics, not considered here, would be the Kawasaki scheme in which up spins diffuse between lattice sites \cite{kawasaki}. 

 All the simulations described here use a lattice of size $L=65$, and coupling constant $J = 0.65 k_BT$. We apply an external magnetic field $h = 0.05 k_BT$ in all simulations. Our coupling constant $J$ is larger than the  critical coupling
$J_c \approx 0.44 k_BT$ \cite{onsager}. Considering the system in the absence of shear, the thermodynamically stable state is ferromagnetic, with net positive magnetisation, meaning that
the system tends to have the majority of its spins in  the up state ($\sigma = +1$). The alternative ferromagnetic state with net negative magnetisation (most spins in the down state) is metastable, and if initiated in a predominantly down state, the system will remain in that state for a significant time before undergoing a nucleation transition to the thermodynamically stable up state \cite{barkema}. We are interested in the rates and pathways for this transition. In the absence of shear, this system is identical to that investigated by Sear \cite{sear} and previously by some of us \cite{barrier}, except that we now use a larger box size, since we have found that the  nucleation rate is more sensitive to
system size in the presence of shear. For the shear rates used in this study, $L=65$ is large enough to ensure that our computed nucleation rate is independent of $L$.   The free energy barrier  in the absence of shear is $\approx 22 k_BT$, so that  we are working at moderate supersaturation. We therefore expect nucleation to proceed via the growth of a single large cluster of up spins. 

We apply  shear to the system using a method similar  to that of Cirillo {\em{et al}} \cite{cirillo}. After each Monte Carlo cycle, we make $M_s \times L$ attempts to shear the system ($M_s$ is the number of attempted shear steps per row per MC cycle). In each attempt, we carry out a shear step with probability $P_s$. A shear step consists of choosing a row $j_s$ at random, and shifting all lattice sites with $j > j_s$ to the right by one lattice site. The net result is that row $j$ is shifted to the right by on average $j M_s P_s$ lattice sites per Monte Carlo cycle. The shear rate  is thus given by $\dot{\gamma} = M_s P_s$. We have verified that the choice of $M_s$ and $P_s$ does not matter for a given product $\dot{\gamma}$. We note that care must be taken to maintain the correct identity of the neighbour sites in the periodic image lattices above and below the simulation box - after a shifting move, the identity of these neighbours is changed. Our method for achieving this is described in Appendix  \ref{app:shear}. This algorithm imposes, on average, a linear velocity field on the underlying lattice. In a real physical system, the velocity field is not imposed externally but emerges as a consequence of the internal dynamics of the system, so that the shear algorithm used here is somewhat artificial. However, our purpose here is to investigate the effects of a linear velocity field on the system; furthermore, this algorithm has the advantages of being simple to implement and homogeneous across the simulation box.

\subsection*{Forward Flux Sampling}

We have used  the Forward Flux Sampling (FFS) method \cite{FFS,FFS2,FFS3} to calculate nucleation rates and transition paths for the formation of the steady state with predominantly up spins (the up state), from the steady state with predominantly down spins (the down state). This rare event sampling method uses a series of interfaces in phase space between the initial and final states to force the system from the initial state $A$ to the final state $B$ in a ratchet-like manner. An order parameter $\lambda(x)$ is defined (where $x$ represents the phase space co-ordinates), such that the system is in state $A$
if $\lambda(x) < \lambda_0$, and it is in state $B$ if $\lambda(x)
> \lambda_n$, while a series of non-intersecting planes (interfaces) $\lambda_i$ ($0 < i < n$) lie between states $A$ and $B$, such that any path from A to B must cross each interface, without reaching $\lambda_{i+1}$ before $\lambda_i$. Provided enough configurations are obtained at the first interface to ensure good sampling, the choice of order parameter $\lambda(x)$ should not affect the calculated rate constant or transition paths ({\em{i.e.}} $\lambda(x)$ need not be the true reaction co-ordinate) - although it may affect the computational efficiency of the method. 

Full details of the FFS method are given in Refs \cite{FFS,FFS2}, and a detailed analysis of its computational efficiency is given in Ref \cite{FFS3}. Briefly,  the transition rate  $I$ from A to B is decomposed as \cite{vanerp1,vanerp2}:
\begin{equation}\label{eq:ffs1}
I=\bar{\Phi}_{\rm{A},n}=\bar{\Phi}_{\rm{A},0}P(\lambda_n|\lambda_0)=\bar{\Phi}_{{\rm A},0}\prod_{i=0}^{n-1}P(\lambda_{i+1}|\lambda_i).
\end{equation}
where  $\bar{\Phi}_{\rm{A},n}$ is the average flux of trajectories crossing from $A$ to $B$, $\bar{\Phi}_{{\rm A},0}$ is the average flux of trajectories crossing $\lambda_0$ in the direction of B, $P(\lambda_n|\lambda_0)$ is the
probability that a trajectory that crosses $\lambda_0$ in the direction of B will eventually reach B before returning to A, and $P(\lambda_{i+1}|\lambda_i)$ is the probability that a trajectory which reaches  $\lambda_i$, having come from A, will reach
$\lambda_{i+1}$ before returning to A. The flux $\bar{\Phi}_{{\rm A},0}$ is computed using a simulation in the A state, during which configurations corresponding to crossings of the first interface $\lambda_0$ coming from A are also stored. This collection of configurations is then used to initiate trial runs which either reach the next interface $\lambda_1$ or go back to  $\lambda_0$, generating an estimate for the conditional probability $P(\lambda_{1}|\lambda_0)$ as well as a new collection of configurations at $\lambda_1$; the trial run procedure is then iterated until $B$ is reached. The rate constant $I$ is then obtained from Eq.(\ref{eq:ffs1}), and a correctly weighted collection of trajectories from $A$ to $B$ is obtained by  tracing trial runs that eventually arrive at $\lambda_n$, via successive interfaces back to $A$.  In practice, rather than storing all configurations for all trial runs during the FFS procedure, it is sufficient to store the initial collection of configurations at $\lambda_0$, together with limited information about each configuration in each collection at intermediate interfaces $\lambda_i$, indicating its ``parent'' configuration in the collection at the previous interface  $\lambda_{i-1}$, and the value of the random number seed used to initiate the relevant trial run. In this way, transition paths can easily be reconstructed after the FFS sampling procedure, without the need for excessive data storage.  In an earlier study, we have shown that FFS correctly reproduces the nucleation behaviour of a two dimensional Ising model in the absence of shear \cite{barrier}.  

For the simulations described in this paper, the parameter $\lambda$ was defined as the total number of up spins in the simulation box. This is a global order parameter: an alternative might be to use the size of the largest cluster of up spins. For FFS, we do not expect the rate constant or the transition path ensemble to depend on the choice of order parameter (we will later analyse trajectories using the largest cluster size). Others have experienced sampling problems when using global order parameters in FFS \cite{koos}; this was not the case in this work. We used 39 interfaces for our FFS calculations (except in section \ref{sec:enh}), defining the $A$ state at $\lambda < \lambda_0$ where $\lambda_0=25$ up spins,
and the $B$ state at $\lambda > \lambda_n$ where $\lambda_n = 2005$ up spins (the total number of spins in our system being $65 \times 65 = 4225$). The spacing between interfaces varies between 5 and 200 up spins. We collected $1000$ configurations at interface $\lambda_0$ and repeated each FFS calculation 25 times. The number of trials at each interface varied between 1300 and 7000. Our results do not depend on the precise choice of the number or position of the interfaces.

\section{Nucleation rate as a function of shear rate}\label{sec:rate}
\begin{figure}[h!]
\begin{center}
{\rotatebox{0}{{\includegraphics[scale=0.3,clip=true]{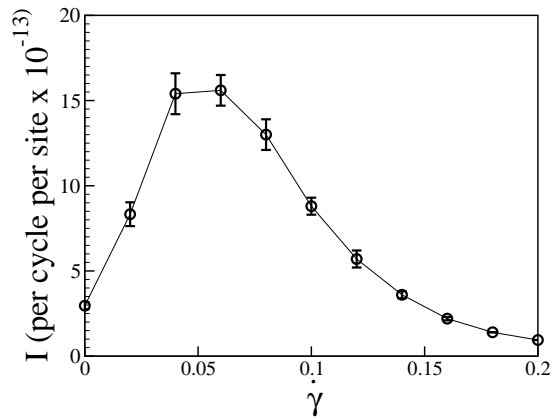}}}}
\caption{Rate of homogeneous nucleation $I$ as a function of shear rate $\dot{\gamma}$ for $h=0.05 k_BT$ and $J = 0.65 k_BT$. 
 \label{fig:Iversusgam}}
\end{center}
\end{figure}

Figure \ref{fig:Iversusgam} shows the rate $I$ of  homogeneous nucleation as a function of the shear rate $\dot{\gamma}$. The nucleation rate shows a striking nonmonotonic dependence on $\dot{\gamma}$.  For low shear rates, $I$ increases apparently linearly with $\dot{\gamma}$, before reaching a maximum around $\dot{\gamma}=0.06$. For shear rates $\dot{\gamma} > 0.06$, $I$ decreases nonlinearly with increasing $\dot{\gamma}$. For $\dot{\gamma}=0$, our result is in good agreement with the value of $3.3 \times 10^{-13}$ per MC cycle per site obtained by Sear \cite{sear}.

\begin{figure}[h!]
\begin{center}
\makebox[20pt][l]{(a)}{\rotatebox{0}{{\includegraphics[scale=0.2,clip=true]{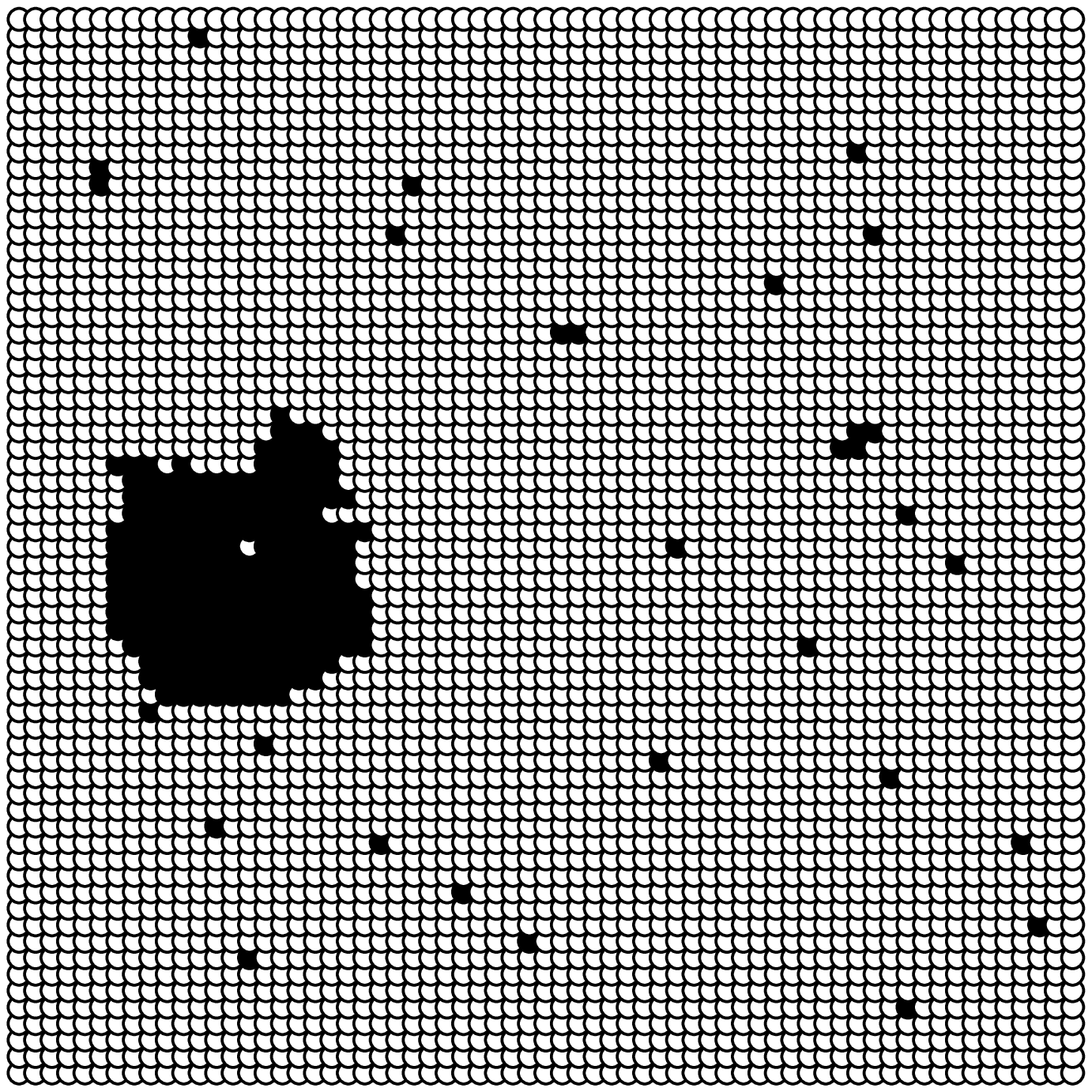}}}}\hspace{1cm}\makebox[20pt][l]{(b)}{\rotatebox{0}{{\includegraphics[scale=0.25,clip=true]{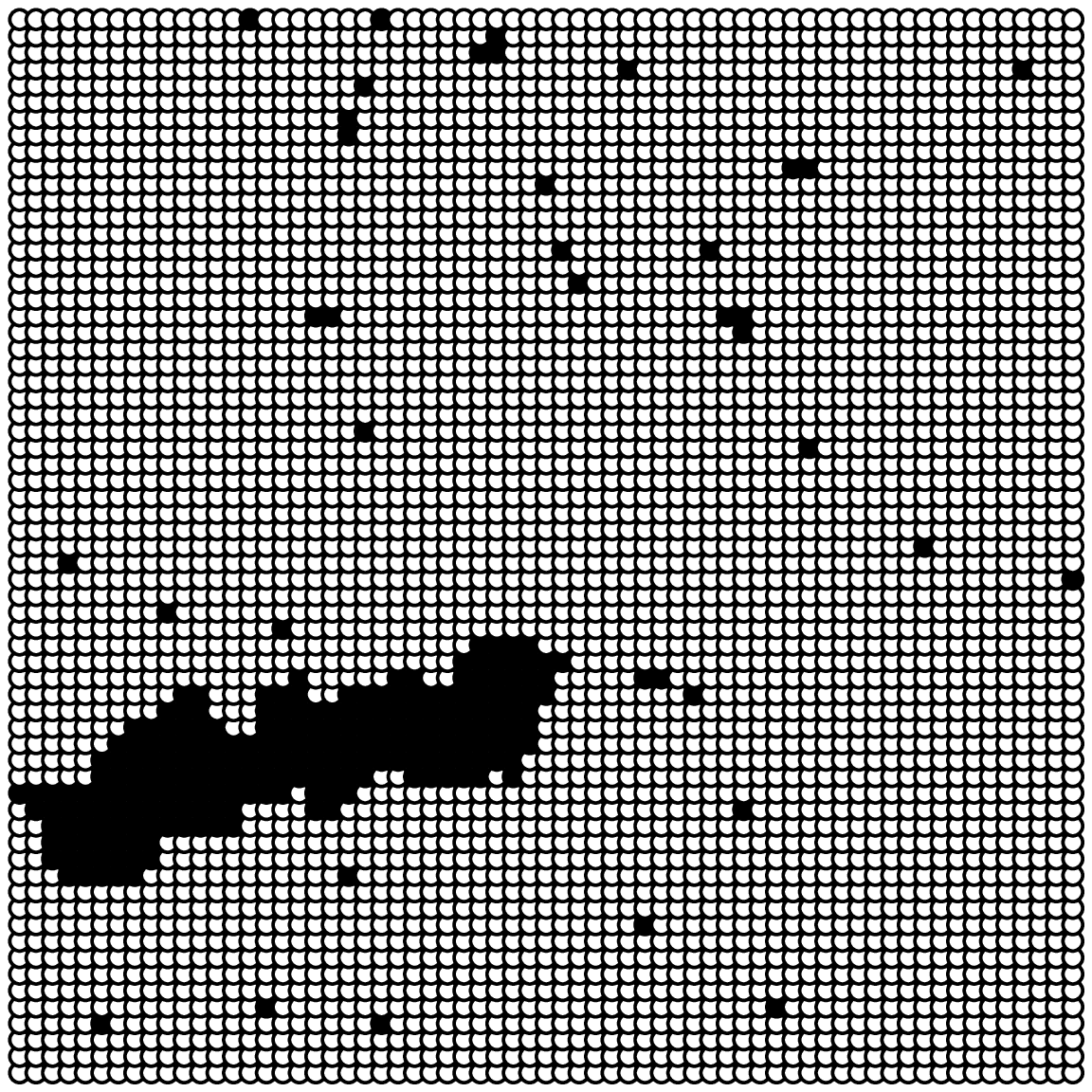}}}}\hspace{1cm}\makebox[20pt][l]{(c)}{\rotatebox{0}{{\includegraphics[scale=0.25,clip=true]{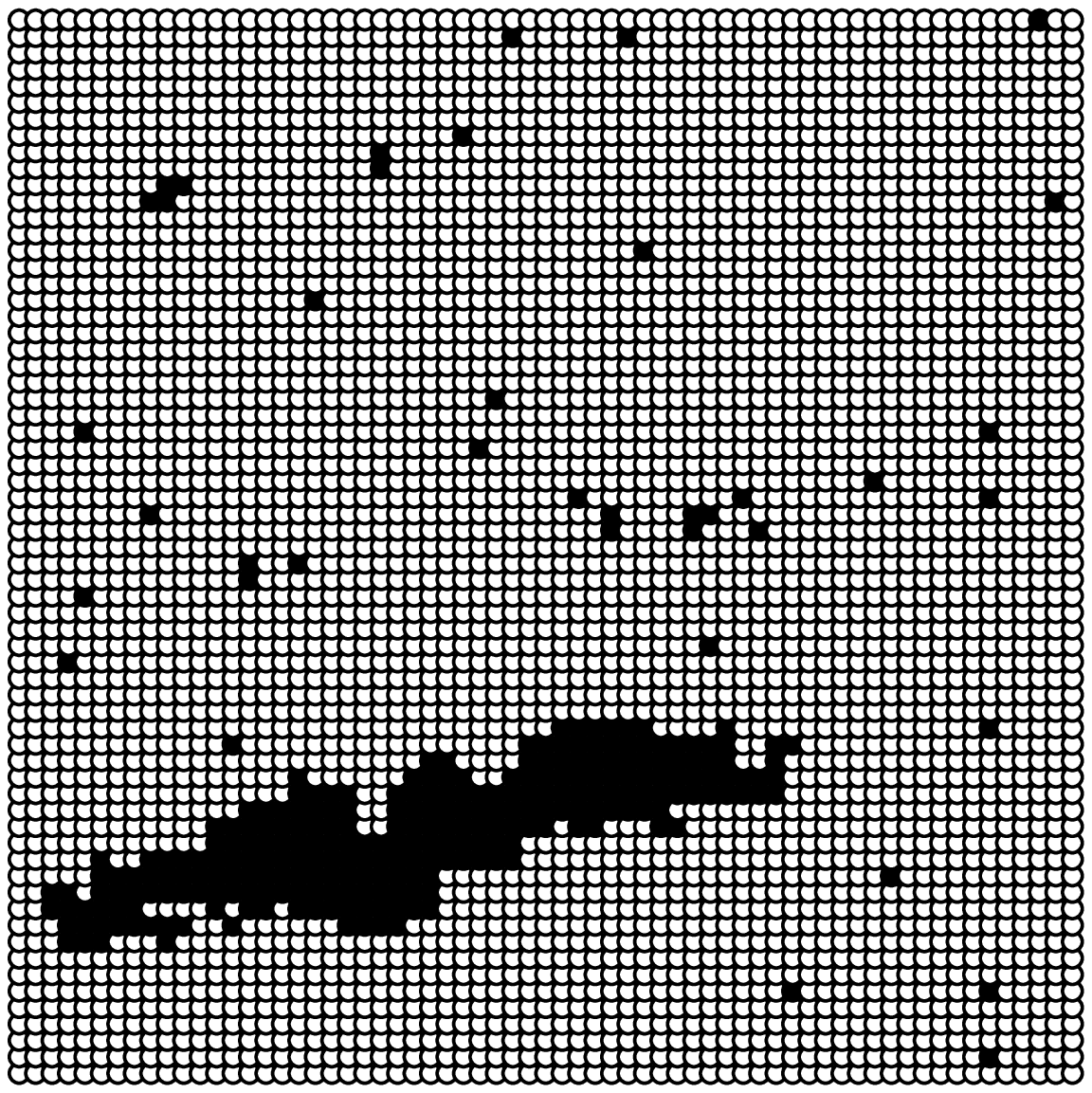}}}}
\caption{{\small Configurations from the transition state ensemble, obtained as described in Appendix \ref{app:tse}  (a): $\dot \gamma = 0.0$,
(b): $\dot \gamma = 0.06$, (c): $\dot \gamma = 0.12$.\label{fig:tse_configs}}}
\end{center}
\end{figure}

Figure \ref{fig:tse_configs} shows representatives of the transition state ensemble (TSE; these are transition path configurations from which a newly initiated trajectory has probability $P_B=0.5$ of reaching $B$ before $A$) for shear rates $\dot \gamma = 0.0$, $\dot \gamma = 0.06$ and $\dot \gamma = 0.12$. It is clear that the shape of the growing cluster is strongly affected by the shear. An observation of the transition paths shows that, for high shear rates, the growing cluster eventually connects with its periodic images to form a horizontal stripe across the box, which then expands vertically to fill the box. This has also been seen for nucleation under shear in the unphysical case of no supersaturation \cite{Venturoli}. In our simulations, stripe formation occurs well beyond the transition state, and does not affect the nucleation rate (since we have verified that $I$ is independent of $L$).

In the following sections, we attempt to elucidate the physical origin of the nonmonotonic dependence of $I$ on $\dot \gamma$ shown in Figure \ref{fig:Iversusgam}. We first consider the origin of the decrease in $I$ with $\dot{\gamma}$ at high $\dot{\gamma}$, and then turn to the mechanisms behind the increase in $I(\dot{\gamma})$ for low $\dot{\gamma}$.

\section{Suppression of nucleation at high shear rate}\label{sec:supp}

We first seek an explanation for the decrease in nucleation rate $I$ with shear rate $\dot{\gamma}$ for $\dot{\gamma} > 0.06$ in Figure \ref{fig:Iversusgam}. Figure \ref{fig:tse_configs} shows that the growing clusters become elongated in the direction of the shear. The extent of this elongation is governed by a balance between the frequency of shear steps and the growth rate of the cluster. It seems intuitive that for high shear rates, the elongation due to the shear will exceed the rate at which the cluster can grow, leading to shear-induced breakup of the growing cluster, and a corresponding decrease in the homogeneous nucleation rate.

\begin{figure}[h!]
\begin{center}
\makebox[20pt][l]{(a)}{\rotatebox{0}{{\includegraphics[scale=0.2,clip=true]{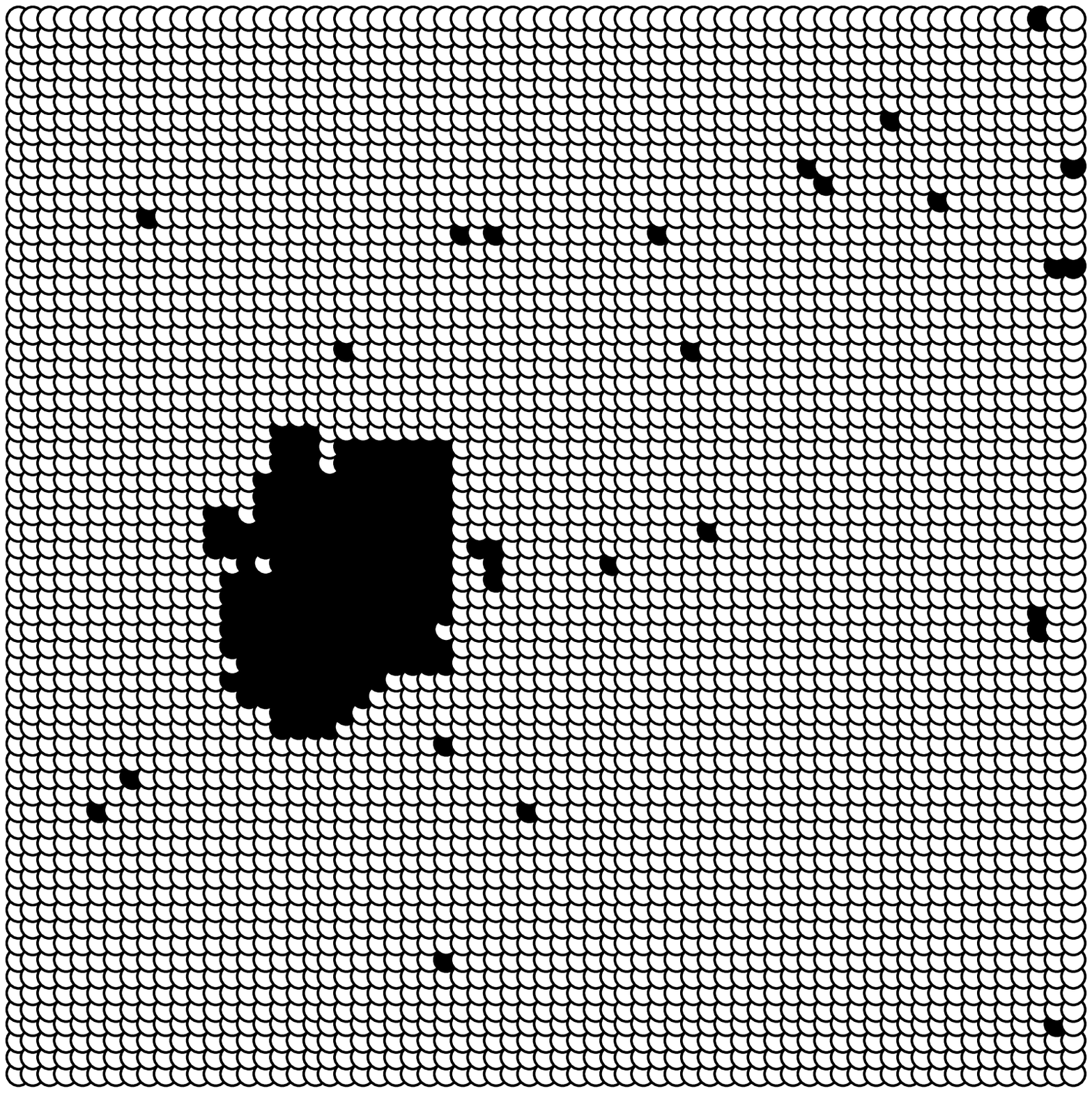}}}}\hspace{1cm}\makebox[20pt][l]{(b)}{\rotatebox{0}{{\includegraphics[scale=0.22,clip=true]{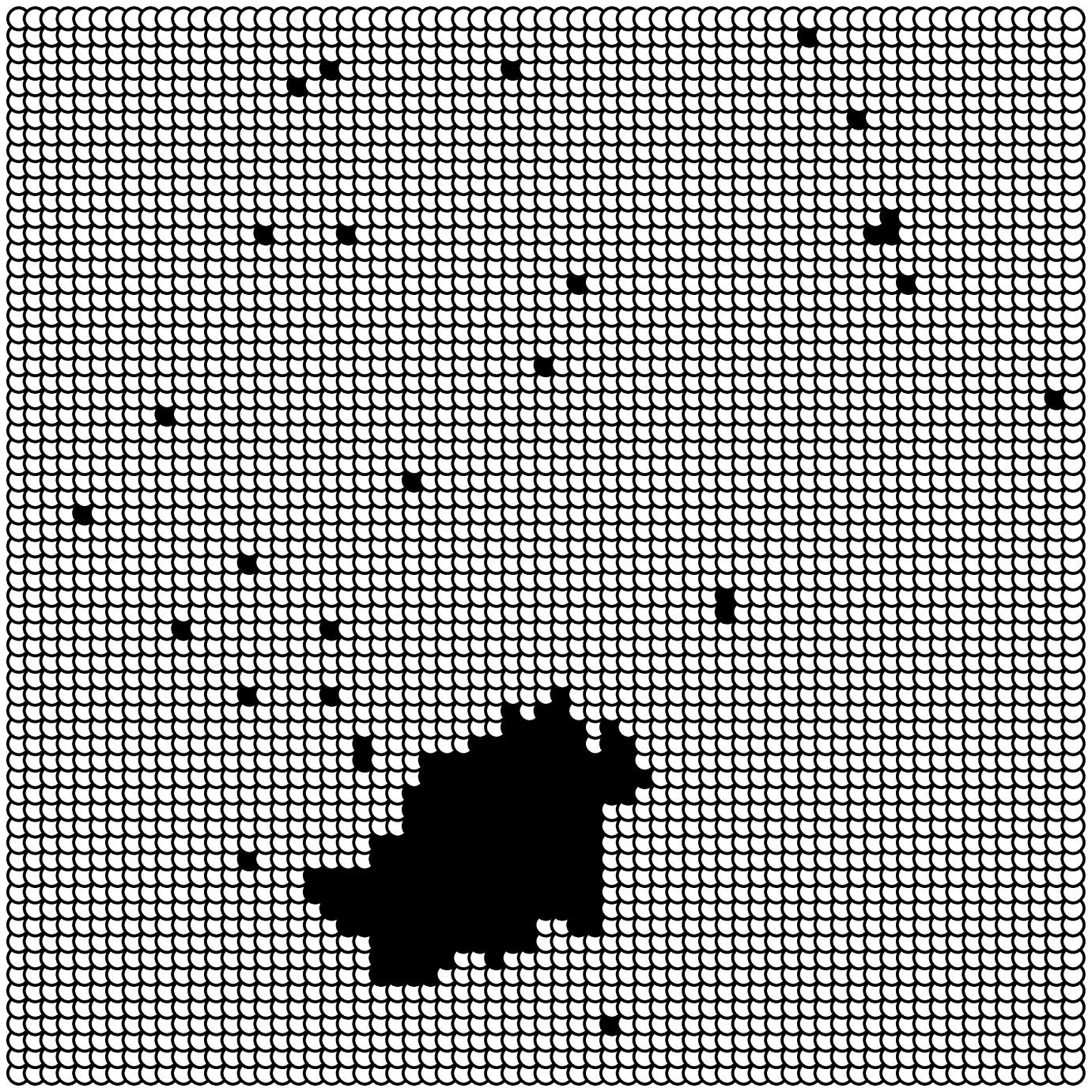}}}}\hspace{1cm}\makebox[20pt][l]{(c)}{\rotatebox{0}{{\includegraphics[scale=0.22,clip=true]{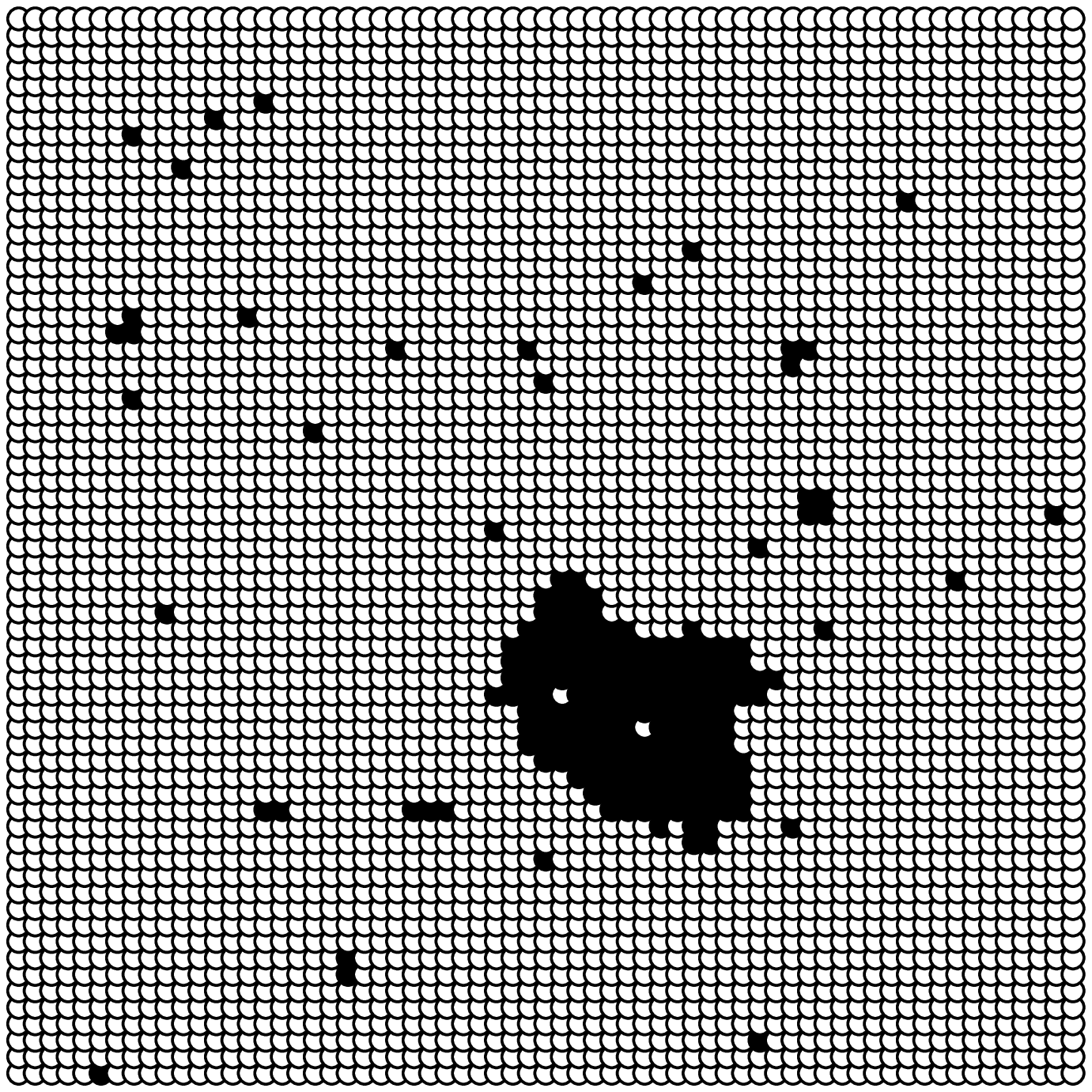}}}}
\caption{{\small Configurations from the transition state ensemble with rattle shear, obtained as described in Appendix \ref{app:tse}  (a): $\dot \gamma = 0.0$,
(b): $\dot \gamma = 0.06$, (c): $\dot \gamma = 0.12$.\label{fig:tse_configs_rattle}}}
\end{center}
\end{figure}

\begin{figure}[h!]
\begin{center}
{\rotatebox{0}{{\includegraphics[scale=0.3,clip=true]{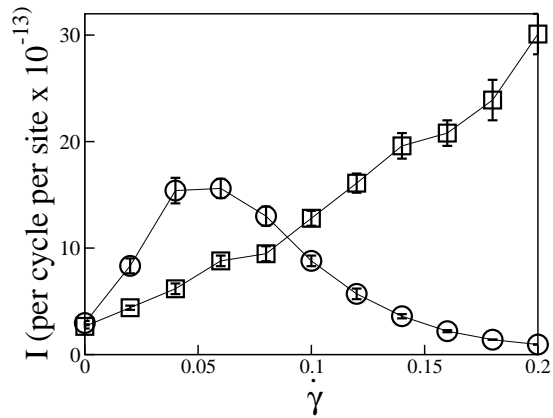}}}}
\caption{$I$ versus $\dot{\gamma}$ for $h=0.05 k_BT$, for regular shear (circles) and rattle shear (squares). The regular shear results are the same as in Figure \ref{fig:Iversusgam}.
 \label{fig:rattle}}
\end{center}
\end{figure}

To test this hypothesis, we performed a set of simulations in which the direction of the shear (to the right or to the left along the $x$ axis) was chosen at random for each shear step. We call this algorithm ``rattle shear''. On average, the system makes as many row shifts to the right as it does to the left, so we do not expect clusters to be elongated by the shear. This is confirmed in Figure \ref{fig:tse_configs_rattle}, which shows that TSE configurations are not noticeably elongated, even for high shear rates. Fig \ref{fig:rattle} shows $I$ versus ${\dot{\gamma}}$ for the rattle shear (squares), as well as for the ``regular'' shear algorithm (circles). As expected, the regime in which $I$ decreases with ${\dot{\gamma}}$ has been abolished for the rattle shear algorithm, at least within this range of ${\dot{\gamma}}$ values. This appears to confirm our hypothesis that the decrease in $I$ for large ${\dot{\gamma}}$ is due to shear-induced elongation of the growing cluster, leading to eventual cluster breakup. 

Figure \ref{fig:rattle} also shows that  nucleation is enhanced less strongly at low shear rates for rattle shear than for the regular shear algorithm. This suggests that multiple shear steps occurring in the same direction are an important feature in the nucleation enhancement mechanism; a topic which is explored further in the next sections.

\section{Enhancement of nucleation at low shear rate}\label{sec:enh}
\begin{figure}[h!]
\begin{center}
{\rotatebox{0}{{\includegraphics[scale=0.45,clip=true]{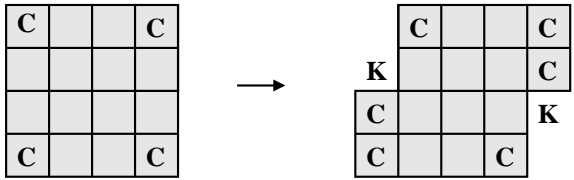}}}}
\caption{A square cluster (left) undergoes one shear step to make a shape with two concave kinks and two additional corner sites. Kink sites are labelled K and corner sites C. 
 \label{fig:kinks}}
\end{center}
\end{figure}

We turn next to the increase in nucleation rate $I$ with shear rate $\dot{\gamma}$ observed in Figure \ref{fig:Iversusgam} for low shear rates, $\dot{\gamma}<0.06$. This behaviour could have (at least) two possible origins. Firstly, the shear algorithm changes the shape of the growing cluster and this may increase its tendency to grow. One way in which this could happen is by increasing the surface roughness of the cluster:  our shear algorithm creates kinks in the growing cluster, and these are favourable sites for further cluster growth. The rate of growth due to spin flips is therefore likely to be enhanced by the shear. This mechanism is illustrated in Figure \ref{fig:kinks}, where two kinks (labelled K) and two corner sites (labelled C) are created by a shear step. Although the corner sites have a tendency to flip to the down state, this is counteracted by the greater tendency (due to the applied field) of the kink sites to flip to the up state. 

\begin{figure}[h!]
\begin{center}
{\rotatebox{0}{{\includegraphics[scale=0.35,clip=true]{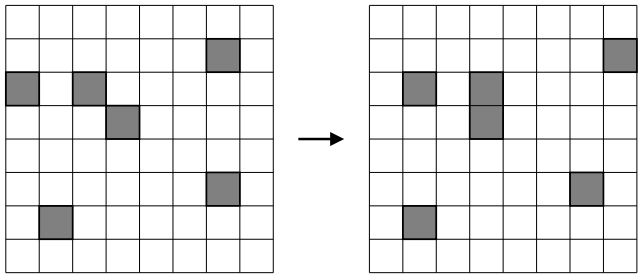}}}}
\caption{The shear algorithm can drive isolated spins together into small clusters. The top three rows are shifted to the right by a shear step, causing two isolated up spins to fuse into a small cluster. 
  \label{fig:clusters}}
\end{center}
\end{figure}

\begin{figure}[h!]
\begin{center}
\makebox[15pt][l]{(a)}{\rotatebox{0}{{\includegraphics[scale=0.3,clip=true]{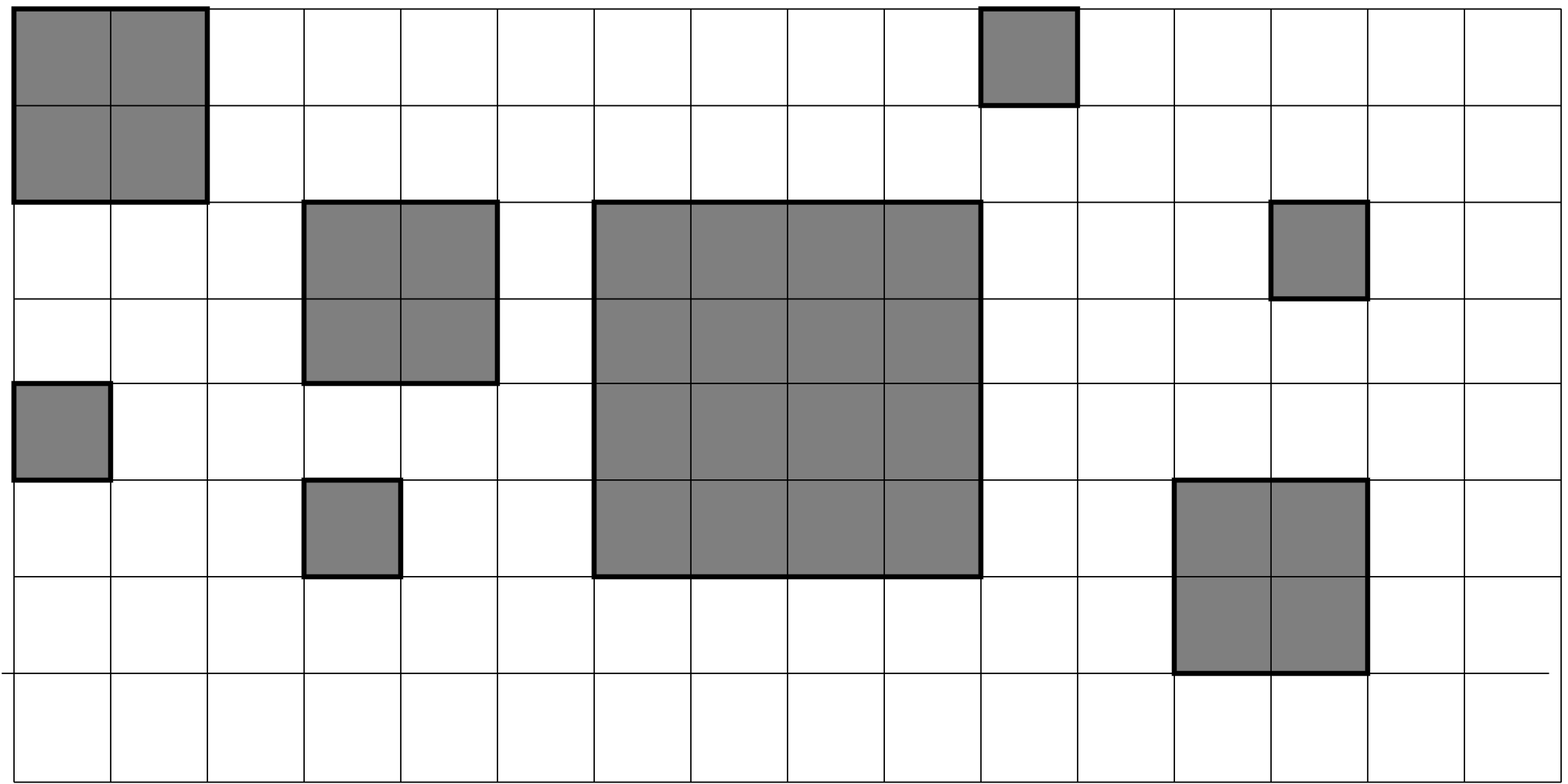}}}\hspace{0.5cm}\makebox[15pt][l]{(b)}\rotatebox{0}{{\includegraphics[scale=0.3,clip=true]{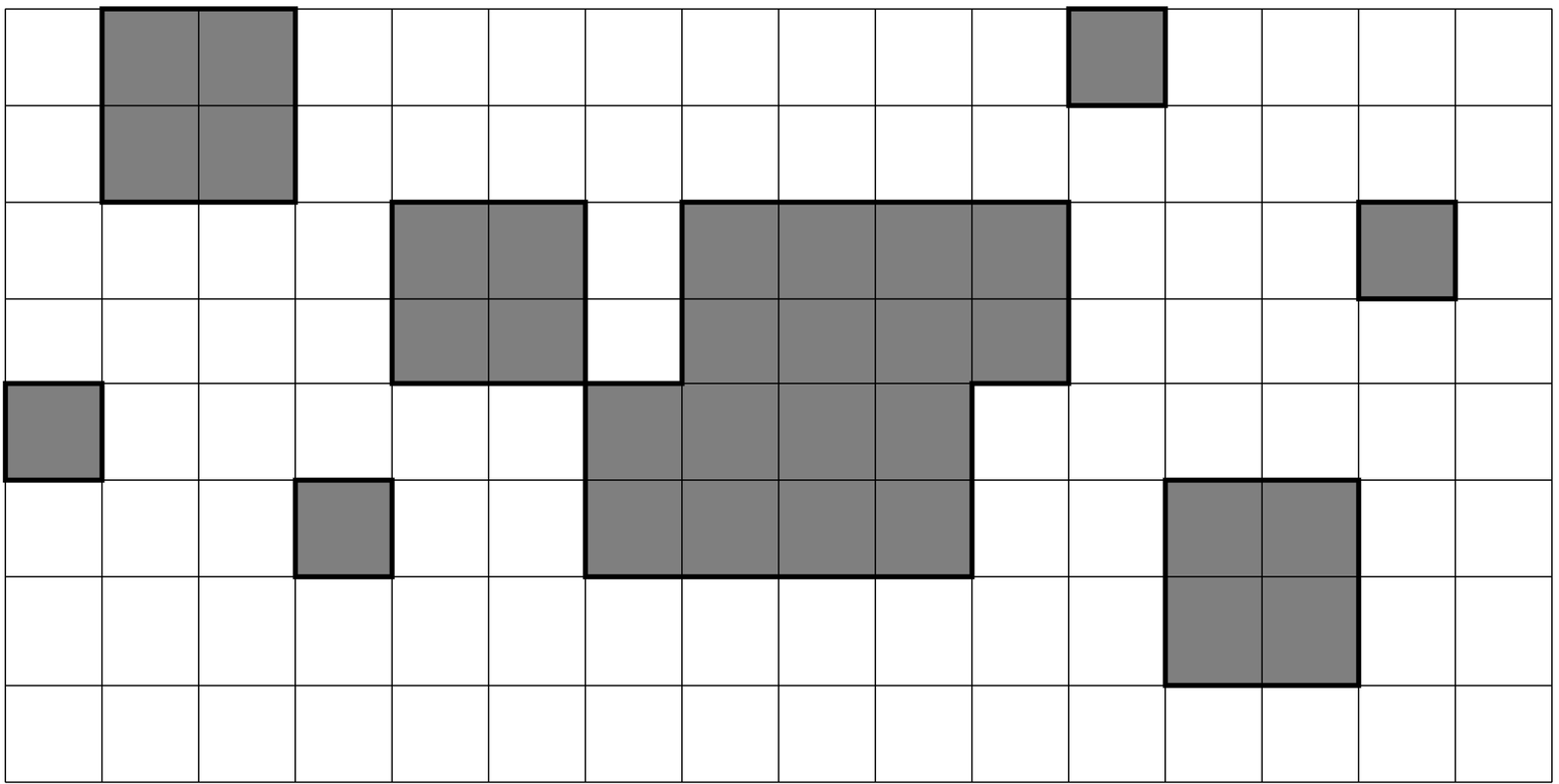}}}\\\vspace{0.5cm}\noindent\makebox[15pt][l]{(c)}\rotatebox{0}{{\includegraphics[scale=0.3,clip=true]{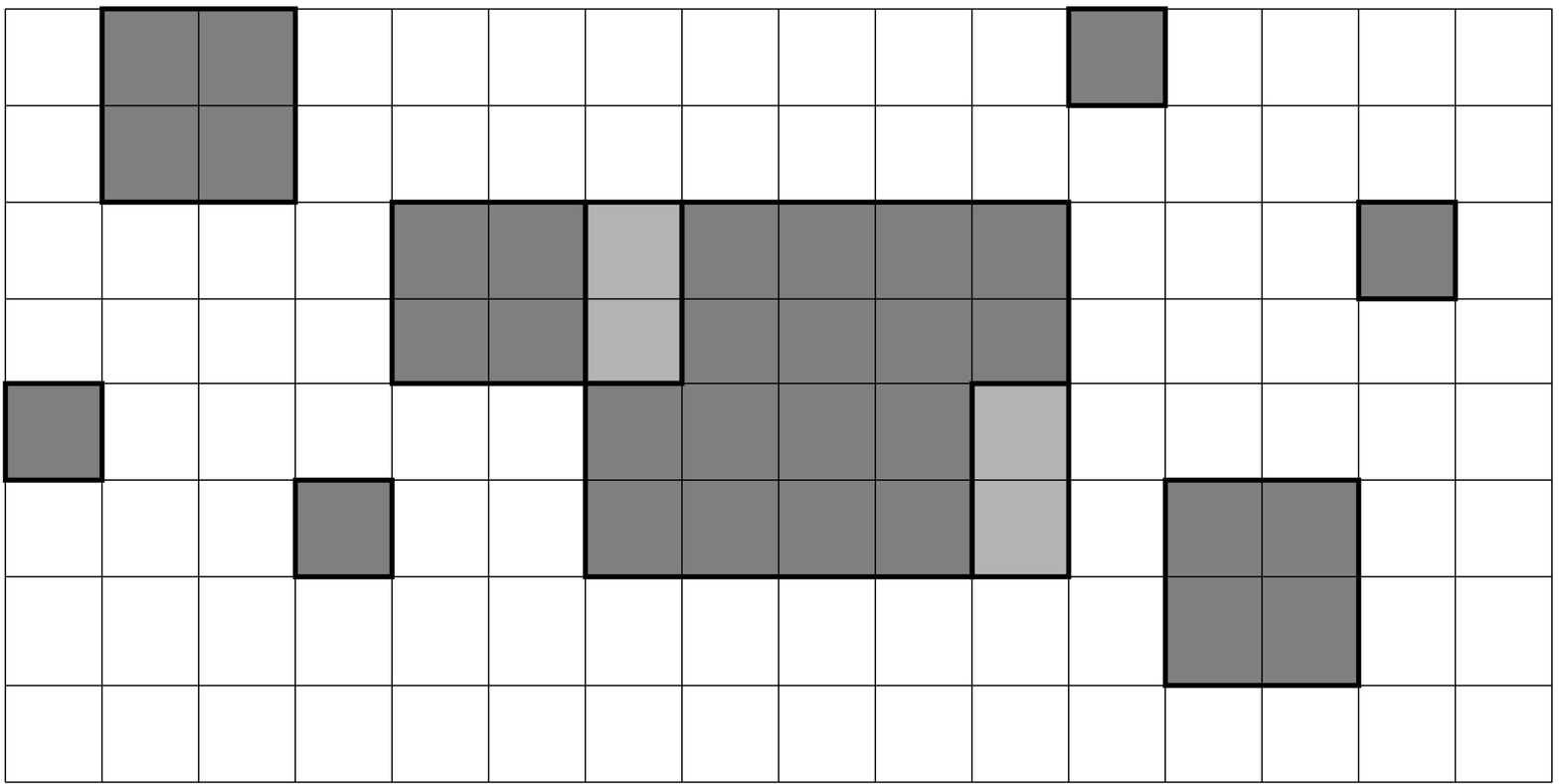}}}\hspace{0.5cm}\makebox[15pt][l]{(d)}\rotatebox{0}{{\includegraphics[scale=0.3,clip=true]{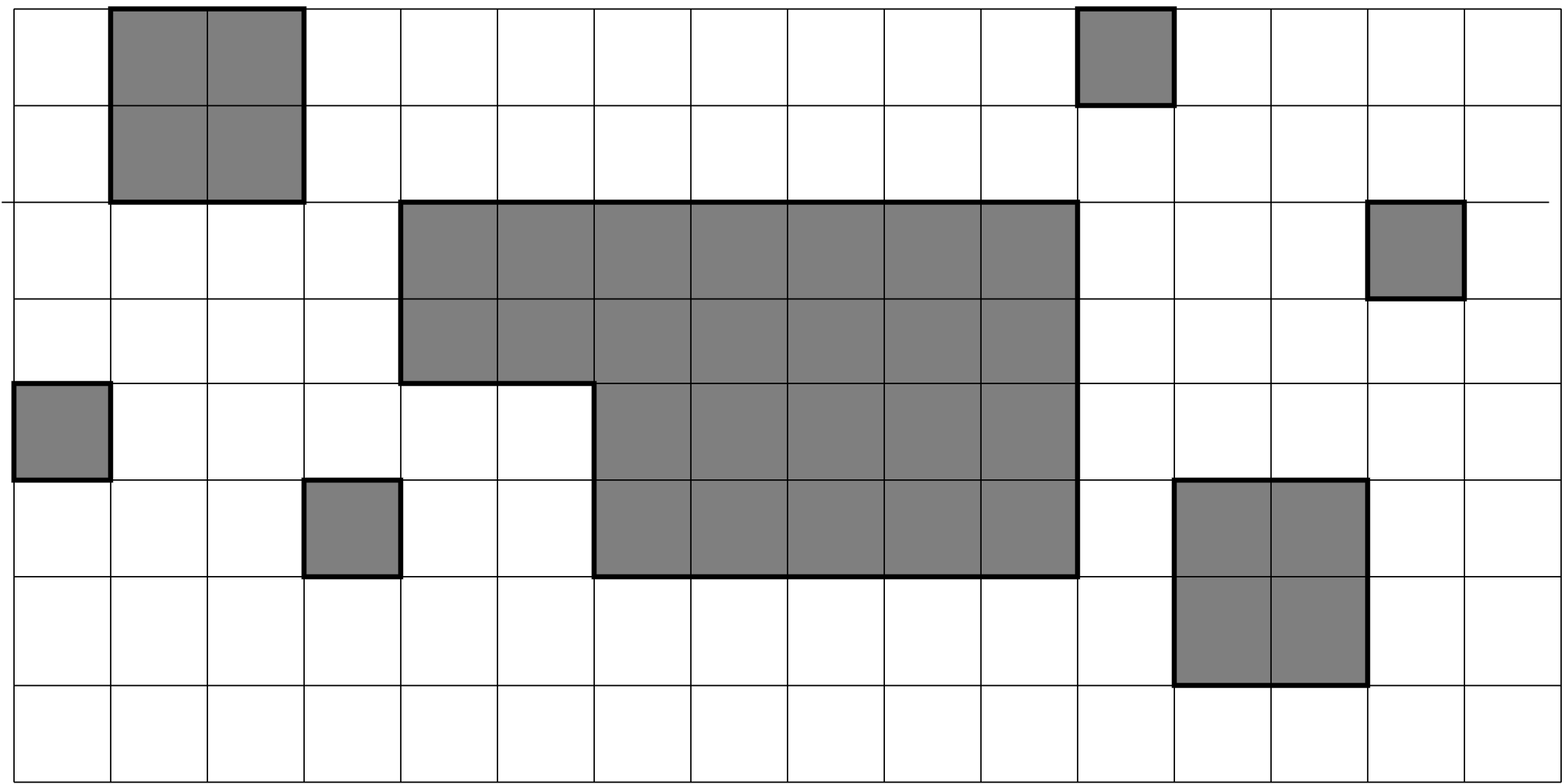}}}}
\caption{The shear algorithm can drive isolated spins and small clusters towards the largest cluster. (a): Initially, a large square cluster is surrounded by 4 isolated ``up'' spins and 3 smaller clusters (b): A shear step occurs and the top half of the simulation box is shifted by one lattice space to the right (c): The cluster grows in the $\pm x$ directions via Metropolis spin flips enhanced by the concave kinks created by the shear (d): In the resulting configuration, the largest cluster has coalesced with one smaller cluster and has also become one lattice space closer to the small cluster at the bottom right.
  \label{fig:coalescence}}
\end{center}
\end{figure}

A second possible mechanism for the increase in $I$ with  $\dot{\gamma}$ is shear-induced coalescence between isolated up spins or small clusters and the growing nucleus. The shear algorithm is expected to drive together isolated up spins, causing an increased abundance of small clusters in the system, as illustrated in Figure \ref{fig:clusters}. These may then coalesce with the largest cluster.  Moreover, isolated spins and small clusters can also be driven towards the largest cluster by the shear algorithm. This is illustrated in Figure \ref{fig:coalescence}. A shear step creates a kink in the largest cluster (Fig \ref{fig:coalescence} a $\to$ b). This is a favourable site for growth, which tends to fill in the clefts created by the kink - with the result that the cluster grows preferentially in the x direction (Fig \ref{fig:coalescence} c). This growth reduces the gap between the largest cluster and surrounding clusters in the $\pm x$ directions (Fig \ref{fig:coalescence} d). Alternatively, multiple shear steps occurring at the second row from the top in Figure \ref{fig:coalescence}a would shift the small cluster at the top left of the box towards the largest cluster, eventually resulting in coalescence, without the need for any growth of the largest cluster. Figures \ref{fig:clusters} and \ref{fig:coalescence} demonstrate that our shear algorithm can promote growth of the largest cluster by coalescence, even though  the system dynamics consists only of Metropolis spin flips and shear steps (diffusion of spins is not modelled).

\section{Analysis of cluster growth}\label{sec:growth}

\begin{figure}[h!]
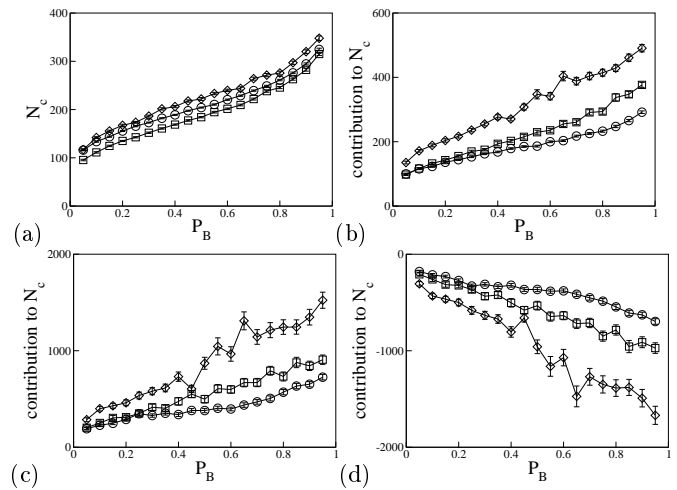

\begin{center}
\makebox[5pt][l]{(a)}{\rotatebox{0}{{\includegraphics[scale=0.17,clip=true]{total_growth.eps}}}}\makebox[5pt][l]{(b)}{\rotatebox{0}{{\includegraphics[scale=0.17,clip=true]{flips.eps}}}}\\\makebox[5pt][l]{(c)}{\rotatebox{0}{{\includegraphics[scale=0.17,clip=true]{coal.eps}}}}\makebox[5pt][l]{(d)}{\rotatebox{0}{{\includegraphics[scale=0.17,clip=true]{break.eps}}}}
\caption{{\small (a): Largest cluster size $N_c$ plotted as a function of the committor $P_B$, averaged over 25 transition paths. (b): Contribution of single spin flips ({\em{i.e.}} spin flips where $\Delta N_c = \pm 1$) to $N_c$. (c) Contribution of coalescence events ({\em{i.e.}} spin flips where $\Delta N_c > 1$) to $N_c$. (d)  Contribution of breakup events ({\em{i.e.}} spin flips where $\Delta N_c < -1$) to $N_c$. In all plots, circles represent results for $\dot \gamma = 0.0$, squares for $\dot \gamma = 0.06$ and diamonds for $\dot \gamma = 0.12$. Note the different scales on the $N_c$ axis. \label{fig:growth}}}
\end{center}
\end{figure}

To elucidate the role of enhanced cluster growth and coalescence in shear-enhanced nucleation, we first analysed the ensemble of transition paths generated by the FFS simulations at shear rates $\dot{\gamma}=0.0$, $\dot{\gamma}=0.06$ and $\dot{\gamma}=0.12$. For each shear rate, we computed the contributions to the growth of the largest cluster of ``single spin flip growth'' (events in which the size of the largest cluster changes by $\pm 1$ spin) and of coalescence (events in which the size of the largest cluster increases by more than one spin). We also measured the contribution of cluster breakup events, in which the size of the largest cluster decreases by more than one spin (an ``event'' here refers to a single attempted spin flip in our Metropolis Monte Carlo scheme, or a shear step). The results are plotted as an average over the transition path ensemble, as a function of the committor $P_B$, in Figure \ref{fig:growth}. The committor function $P_B(x)$ is the probability that a trajectory initiated from a configuration $x$ will reach the final state $B$ before the initial state $A$ - this provides a convenient measure for the progress of the transition.

Figure \ref{fig:growth}a shows that, for all shear rates, the largest cluster increases in size as the transition progresses in a rather similar manner, although the shear causes a slight difference in cluster size: $N_c(\dot{\gamma}=0.06) < N_c(\dot{\gamma}=0.0) < N_c(\dot{\gamma}=0.12)$. However, Figure \ref{fig:growth}b, c, and d show that despite this apparent similarity, the contributions of single spin flips, coalescence and breakup events to cluster growth are all strongly affected by the shear. Both single spin flip growth and coalescence are enhanced for $\dot{\gamma}=0.06$ compared to the zero shear case, and strongly enhanced for $\dot{\gamma}=0.12$. However, this is balanced by a strong increase in the negative contribution of cluster breakup events in the presence of shear. This suggests that both the mechanisms outlined above (shear-enhanced single spin flip growth and shear-enhanced coalescence), as well as cluster breakup, are likely to play a significant role in the nucleation mechanism in the presence of shear.

\section{A modified shear algorithm}\label{sec:mod}

We would like to quantify the contributions of shear-enhanced cluster growth and coalescence to the increase in nucleation rate with shear rate shown in Figure \ref{fig:Iversusgam}. To do this, we devised a modified shear algorithm in which shear-induced coalescence is largely eliminated. This is the same as the regular shear algorithm (see  section \ref{sec:sim}), except that after each shear step, spins in the rows immediately above and below the ``break point'' are allowed to equilibrate with no shear for $N_{eq}$ Metropolis Monte Carlo steps. During this equilibration, all spins in the largest cluster and its immediate neighbours are held fixed, even if they lie in the rows mentioned above. The effect of this equilibration is that the largest cluster (and its immediate neighbours) is sheared as normal, while the surrounding ``bath'' of spins is maintained close to equilibrium (there may still be small clusters in the bath which have broken off from the largest cluster). It is not necessary to equilibrate all rows after a shear step. Our system has only nearest-neighbour interactions, and the effect of the shear is simply to change the juxtaposition of the two rows adjacent to the break point; all other rows remain unchanged during a shear step.  We have used $N_{eq}=5$ equilibration steps after each shear step (although in fact we find that $N_{eq}=1$ is sufficient). 

\begin{figure}[h!]
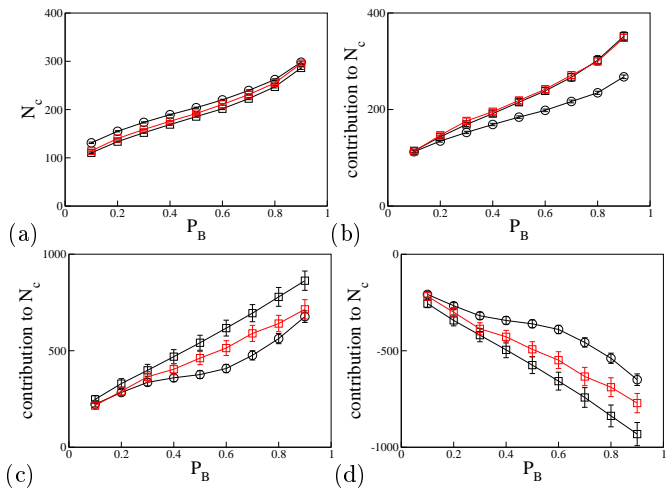

\begin{center}
\makebox[5pt][l]{(a)}{\rotatebox{0}{{\includegraphics[scale=0.17,clip=true]{total_growth_compare.eps}}}}\makebox[5pt][l]{(b)}{\rotatebox{0}{{\includegraphics[scale=0.17,clip=true]{flips_compare.eps}}}}\\\makebox[5pt][l]{(c)}{\rotatebox{0}{{\includegraphics[scale=0.17,clip=true]{coal_compare.eps}}}}\makebox[5pt][l]{(d)}{\rotatebox{0}{{\includegraphics[scale=0.17,clip=true]{break_compare.eps}}}}
\caption{{\small Contributions to the largest cluster size as in Figure \ref{fig:growth}, for the modified and regular shear algorithms for  $\dot \gamma = 0.06$. Red squares: modified shear algorithm; black squares: regular shear algorithm (both for $\dot \gamma = 0.06$). Results are compared to those for the regular shear algorithm in the absence of shear, shown by the black circles. (a): Largest cluster size $N_c$. (b): Contribution of single spin flips ({\em{i.e.}} spin flips where $\Delta N_c = \pm 1$) to $N_c$. (c) Contribution of coalescence events ({\em{i.e.}} spin flips where $\Delta N_c > 1$) to $N_c$. (d)  Contribution of breakup events ({\em{i.e.}} spin flips where $\Delta N_c < -1$) to $N_c$.  Note the different scales on the $N_c$ axis. \label{fig:growth_compare}}}
\end{center}
\end{figure}

We expect this modified algorithm significantly to reduce shear-induced cluster coalescence. The  shear cannot form small clusters or drive surrounding spins and clusters towards the largest cluster, as illustrated in Figure \ref{fig:coalescence}. Coalescence events in which the shear directly merges clusters, as illustrated in  Figure \ref{fig:clusters}, may still occur, however, as may events where the largest cluster is broken up by the shear and subsequently re-coalesces. This type of modified shear algorithm might also be useful for investigating the effects of coalescence for more complex systems with long-range interactions or off-lattice particles: however, in these cases it would be necessary to equilibrate the whole bath after every shear step (or from time to time in the case of continuous shear). If the nucleation mechanism proceeds via the coalescence of multiple large clusters (at large supersaturation), this approach is unlikely to be useful. However, our simulations are at moderate supersaturation so that we believe nucleation occurs via the growth of a single large cluster.

Figure \ref{fig:growth_compare} shows the contributions to cluster growth of single spin flips, coalescence and breakup (as in Figure \ref{fig:growth}), for the modified shear algorithm, compared to the regular algorithm, for $\dot{\gamma}=0.06$ (moderate shear). Transition paths were extracted from a FFS calculations: for the modified algorithm, we used 32 interfaces and $\lambda_B=805$, as the algorithm is computationally expensive. We verified that this change of FFS parameters has no effect on the computed rate constant, since by  $\lambda=805$ the paths are completely committed to nucleation.

As expected, single spin flip growth is enhanced by shear for the modified algorithm in the same way as for the regular algorithm (in Figure \ref{fig:growth_compare}b, the data for the modified algorithm with $\dot{\gamma}=0.06$ overly those for the regular algorithm with $\dot{\gamma}=0.06$). Turning to Figure \ref{fig:growth_compare}c, we see that the coalescence contribution is significantly reduced for the modified algorithm (red squares) compared to the regular algorithm (black squares). However, some shear enhancement of coalescence still remains in the modified algorithm, since the results for the modified algorithm with $\dot{\gamma}=0.06$ (red squares) do not coincide with those for the regular algorithm in the absence of shear (black circles). We speculate that these remaining coalescence events involve clusters breaking up and then re-coalescing (these events are not suppressed by our modified algorithm). Another possibility may be that the shear changes the largest cluster in such a way that it becomes more prone to attaching other clusters, even in an equilibrium bath. Figure \ref{fig:growth_compare}d shows the contribution of cluster breakup. Cluster breakup is also partially reduced in the modified algorithm compared to the regular algorithm. We expect the largest cluster to be stretched by the shear in the same way for both algorithms - so we might expect the contribution of cluster breakup to be the same for both algorithms. However, some coalescence events result immediately in breakup, if the incoming small cluster is not well attached to the largest cluster. Suppressing these coalescence events in our modified algorithm may therefore also decrease the rate of breakup events.

\begin{figure}[h!]
\begin{center}
{\rotatebox{0}{{\includegraphics[scale=0.3,clip=true]{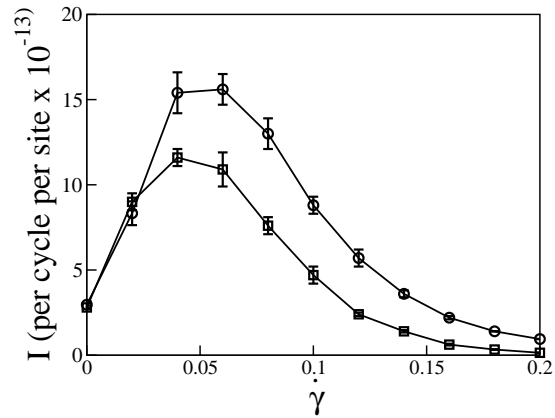}}}}
\caption{$I$ versus $\dot{\gamma}$ for $h=0.05 k_BT$, for ``regular'' shear (circles) and the modified shear algorithm (squares). The ``regular'' shear results are the same as in Figure \ref{fig:Iversusgam}.
 \label{fig:nocoal}}
\end{center}
\end{figure}

Figure \ref{fig:nocoal} shows the nucleation rate $I$ as a function of the shear rate ${\dot{\gamma}}$ for the modified algorithm, as well as the original results from Figure \ref{fig:Iversusgam}.    The modified algorithm gives the same qualitative behaviour for $I({\dot{\gamma}})$, but the nucleation rate is reduced by a factor which increases with the shear rate. These results suggest that shear-induced interactions between the growing cluster and the surrounding sheared ``bath'' of spins play a significant role in enhancing nucleation for shear rates $\dot{\gamma} > 0.03$.

\section{Analysis of the transition state ensemble}{\label{sec:antse}

To further test the role of cluster coalescence in the nucleation mechanism, we  have analysed the committor function $P_B(x)$. This is the probability that a trajectory initiated from a configuration $x$ will arrive in the $B$ state before the $A$ state. Surfaces in configurational space on which $P_B$ takes a fixed value are known as the isocommittor surfaces. In particular, the transition state surface has  $P_B=0.5$. It is important to be clear about which configurations $x$ we use to define the isocommittor surfaces. Because our system is driven by shear and its dynamics do not obey detailed balance, the path ensembles for the forward and reverse transitions need not lie in the same region of configurational space, so that the transition state surfaces for the forward and reverse transitions need not be the same for a nonequilibrium problem \cite{FFS}. Even for cases where detailed balance is obeyed, it is important to define whether the isocommittor surface is computed using configurations from the path ensemble or using Boltzmann-weighted configurations sampled from the entire phase space. We will study here the transition state surface for the forward paths: {\em{i.e.}} we will analyse the collection of configurations in the forward path ensemble with  $P_B=0.5$. These configurations are members of the transition state ensemble (TSE) for the forward transition. We will carry out this analysis for three different shear rates $\dot{\gamma} = 0.0$, $\dot{\gamma} = 0.06$ and  $\dot{\gamma} = 0.12$. The TSE configurations were extracted from the transition paths as described in Appendix \ref{app:tse}.

Committor analysis can be used to test whether a chosen order parameter $\mu$ is important in the transition mechanism, by computing the correlation between the value of  $\mu$ and the committor value, for configurations in the transition paths. If $\mu$ is found to be strongly correlated with the committor, then it is likely that this order parameter captures (at least some of) the essential physics underlying the transition. For this nucleation problem, we know that the largest cluster size $N_c$ is an order parameter that couples strongly to the committor - large clusters have a greater probability of continuing to grow than small clusters. In Classical Nucleation Theory, it is assumed that $N_c$ is the only important order parameter. However, in the presence of shear, other order parameters, which couple to the shear, must also be important. Since we have postulated that cluster coalescence plays an important role in the nucleation mechanism in the presence of shear, we seek an order parameter  $\mu$ that is coupled to coalescence.

\begin{figure}[h!]
\begin{center}
\makebox[20pt][l]{(a)}{\rotatebox{0}{{\includegraphics[scale=0.2,clip=true]{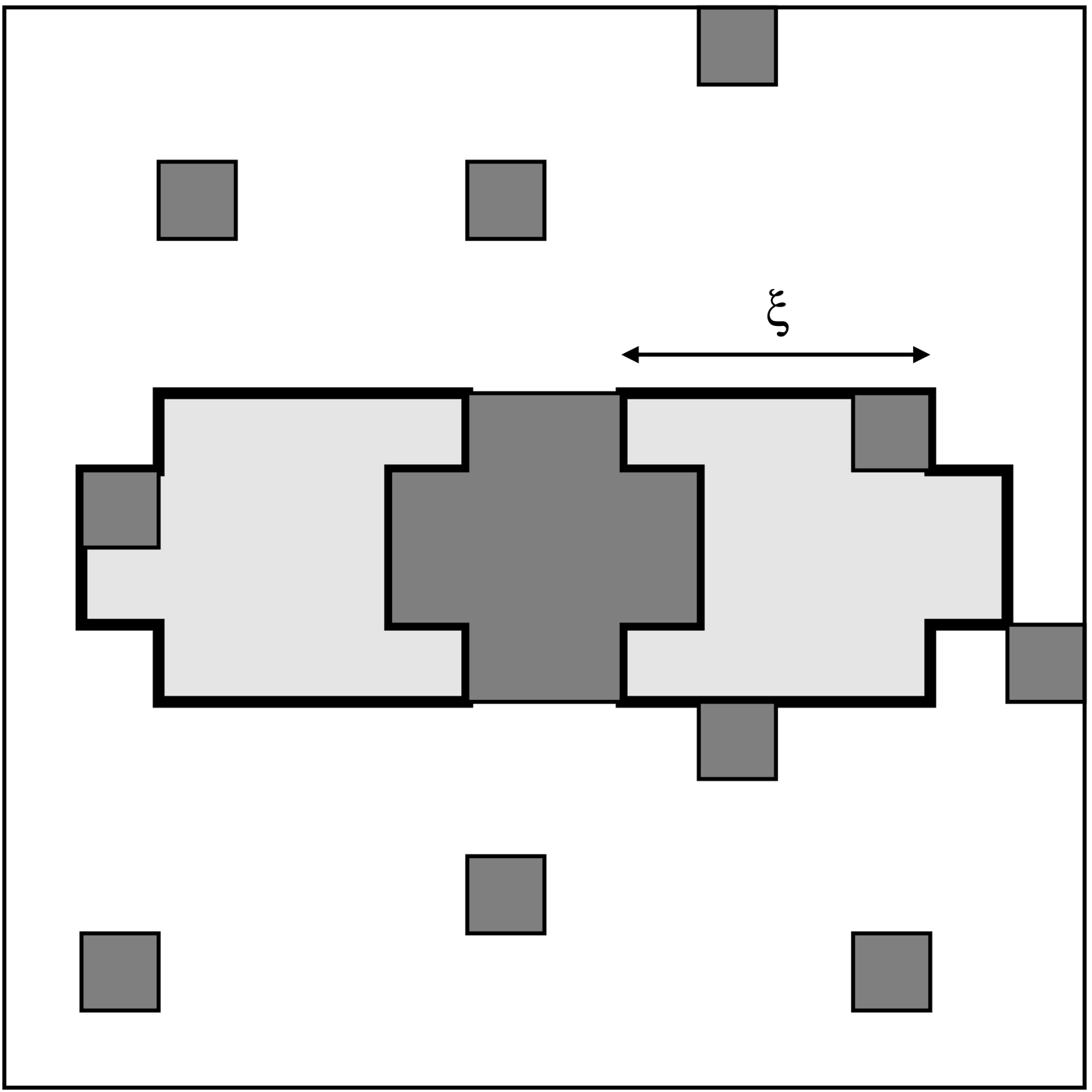}}}\makebox[20pt][l]{(b)}\rotatebox{0}{{\includegraphics[scale=0.2,clip=true]{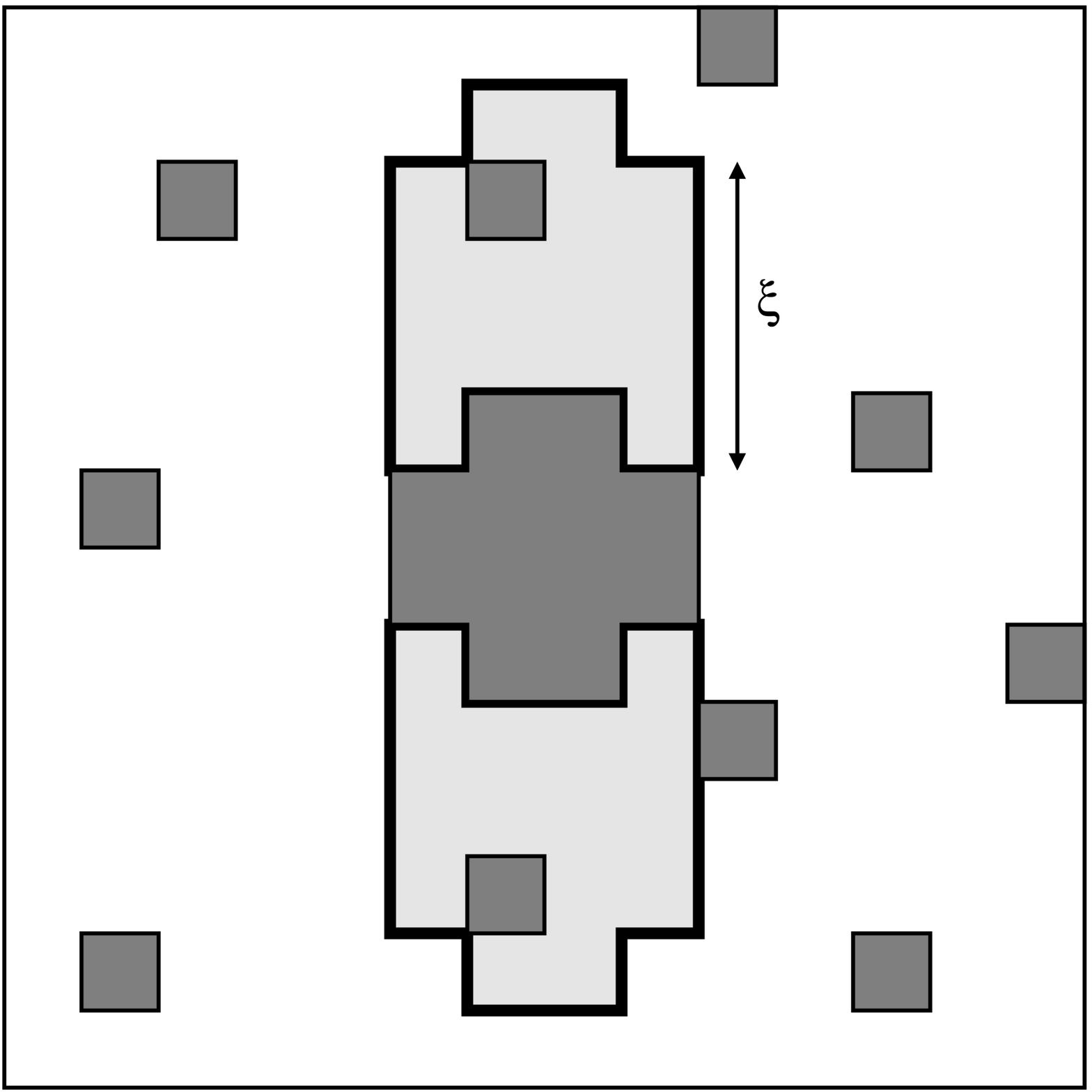}}}}
\caption{Order parameters for local density of ``up'' spins: $\rho_{x}$ is the density of ``up'' spins, not in the largest cluster, in the shaded region in panel (a), while $\rho_{y}$ is the density of ``up'' spins, not in the largest cluster, in the shaded region in panel (b). 
 \label{fig:rldens}}
\end{center}
\end{figure}

Figure \ref{fig:rldens} illustrates one such order parameter: the local density of up spins surrounding the largest cluster. The coalescence mechanism illustrated in Figure  \ref{fig:coalescence} is expected to depend on the density  of up spins in the regions to the right and left of the cluster, but not in the regions directly above and below. We therefore define local densities  $\rho_{x}$ and $\rho_{y}$ as the density of up spins surrounding the largest cluster, located a distance $|x| \le \xi$ or $|y| \le \xi$ from the edge of the largest cluster respectively, where $\xi$ is a cutoff distance. These regions are shown schematically in  Figure \ref{fig:rldens}a and b. If shear-induced coalescence is indeed important in the nucleation mechanism, we expect $\rho_{x}$ but not $\rho_y$ to be correlated with the committor in the presence of shear. Neither $\rho_{x}$ nor $\rho_y$ is expect to correlate with the committor in the absence of shear.

To test whether $\rho_{x}$ and $\rho_y$ are correlated with the committor, we extract TSE configurations (which have $P_B=0.5$) from our transition paths, and look for negative correlation between $\rho_{x}$ or $\rho_y$ and the largest cluster size $N_c$ for these configurations. We know that  $N_c$ is an important order parameter for this transition. If another order parameter $\mu$ is also important then we would expect both $N_c$ and $\mu$ to determine the committor value. Specifically, TSE configurations can achieve $P_B=0.5$ by having a large value of $N_c$ but only a small value of $\mu$, or alternatively a small value of  $N_c$ may be compensated by a large value of $\mu$. This implies that if  $\mu$ is positively correlated with the committor, then  for the TSE configurations, $\mu$ should be negatively correlated with $N_c$.

\begin{figure}[h!]
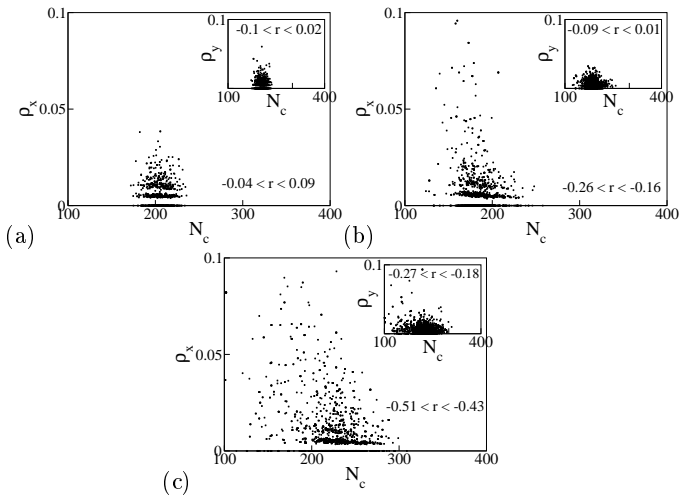

\begin{center}
\makebox[5pt][l]{(a)}{\rotatebox{0}{{\includegraphics[scale=0.17,clip=true]{0.0_localdens.eps}}}\makebox[5pt][l]{(b)}\rotatebox{0}{{\includegraphics[scale=0.17,clip=true]{0.06_localdens.eps}}}\\\makebox[5pt][l]{(c)}\rotatebox{0}{{\includegraphics[scale=0.17,clip=true]{0.12_localdens.eps}}}}
\caption{Scatter plots of $\rho_{x}$ (main panel) or $\rho_{y}$ (inset)  versus size of largest cluster $N_c$ for TSE configurations (defined as configurations in transition paths with $P_B=0.5$), for (a): $\dot{\gamma}=0.0$, (b): $\dot{\gamma}=0.06$ and (c): $\dot{\gamma}=0.12$. Order parameters $\rho_{x}$ and $\rho_y$ were defined as above with $\xi=5$ lattice sites. 95\% confidence intervals for the correlation coefficient (``r value'') between $\rho$ and $N_c$ are given in each panel \cite{clarke}. 
 \label{fig:xdensscatter}}
\end{center}
\end{figure}

Figure \ref{fig:xdensscatter} tests whether this negative correlation exists. The main panels show  scatter plots of $\rho_x$  versus $N_c$ for configurations in the TSE (each configuration is represented by one dot), for $\xi=5$ lattice sites. In the absence of shear ($\dot{\gamma}=0.0$), there is no significant correlation between  $\rho_x$ and $N_c$, but significant negative correlation is observed for $\dot{\gamma}=0.06$ and $\dot{\gamma}=0.12$. The correlation coefficient (``r-value'') is larger for $\dot{\gamma}=0.12$ than for $\dot{\gamma}=0.06$, suggesting that the importance of coalescence increases with shear rate. The insets in Figure \ref{fig:xdensscatter} show the same scatter plots for  $\rho_y$.   No significant correlation is present in the absence of shear (a) or when $\dot{\gamma}=0.06$ (b). This supports our hypothesis that local density in the x but not the y direction is important in the presence of shear. For the highest shear rate, $\dot{\gamma}=0.12$, we do observe significant negative correlation between $\rho_y$ and $N_c$, but the magnitude of the correlation coefficient is much less for $\rho_y$ than for $\rho_x$. Taken together, these results  suggest that the shear-induced coalescence mechanism illustrated in Figures \ref{fig:clusters} and \ref{fig:coalescence} does capture some of the essential physics of the nucleation mechanism. 

Since we believe that shear also enhances ``single spin flip'' cluster growth, we would like to  repeat the above analysis for an order parameter that measures the tendency of the cluster to grow. However, this has proved difficult. Order parameters based on measuring the number of kinks in the perimeter of the largest cluster tend to covary with the size of the largest cluster, so that scatter plots of $\mu$ versus $N_c$ are not an objective measure. These order parameters are also affected by the shape of the cluster, which is affected by the shear. Attempts to measure directly the propensity of the cluster to grow by single spin flips failed to produce significant negative correlation, possibly because this propensity fluctuates greatly during cluster growth.

 \section{Discussion}\label{sec:dis}

In this paper, we have used Forward Flux Sampling (FFS) to calculate rate constants and transition paths for homogeneous nucleation in a sheared two dimensional Ising model. Our results show a striking nonmonotonic dependence of the nucleation rate $I$ on the shear rate ${\dot{\gamma}}$. We have investigated the physical mechanisms underlying this behaviour by analysing transition paths and the transition state ensemble, as well as by comparison to several modified shear algorithms. Our conclusions are that the observed decrease in $I(\dot{\gamma})$ for large $\dot{\gamma}$ is due to shear-mediated breakup of the growing cluster, while the increase in $I(\dot{\gamma})$ for small $\dot{\gamma}$ is due to shear-induced cluster coalescence as well as to shear-enhanced ``single spin flip'' growth of the largest cluster. The contributions of shear-enhanced cluster coalescence and single spin flip growth appear to be of the same order of magnitude.

Our analysis has been strongly influenced by work on (quasi-)equilibrium systems, in which the goal is often to identify the ``reaction coordinate''. This is the collective coordinate which most strongly correlates with the committor for the transition. If the true reaction coordinate could be identified, it is believed that the transition could be coarse-grained into a one-dimensional diffusion problem, over a free energy barrier in the reaction coordinate, in the manner of Classical Nucleation Theory. Several methods have been proposed for identifying the optimal reaction coordinate \cite{ma,peters1,peters2}. We were motivated by this approach in our attempts to find order parameters that correlate negatively with $N_c$ in Section \ref{sec:enh}. However, it is not clear whether the reaction coordinate is such a useful concept for driven systems as it is for equilibrium systems. As discussed in section \ref{sec:antse}, the isocommittor surfaces for the forward and backward transitions are not necessarily the same for driven systems, so the reaction coordinates may be different for the forward and backward processes. Moreover, the reaction coordinate is usually assumed to be a function only of the spatial coordinates of the system - whereas in systems that are not in local equilibrium, dynamical coordinates (momenta) may also be important. Even if a collective coordinate which correlated precisely with the committor could be found, it is not clear whether the complex dynamics of the driven system could then be coarse-grained into a one-dimensional model. Our preliminary investigations suggest that  it may be possible to construct a one-dimensional ``CNT-like'' model for Ising nucleation under shear, along the lines of a toy model proposed by Cerd{\`{a}} {\em{et al}} \cite{cerda,avt}.

The two dimensional Ising model studied in this paper is an idealised system, and the algorithm used here to apply shear is far from realistic. In particular, our dynamics includes no spin transport and our shear algorithm imposes a linear ``velocity gradient'' on the system. In experimental systems, mass transport, hydrodynamic effects, etc, undoubtedly play a role. Effects of mass transport could be included using an Ising model with Kawasaki dynamics \cite{kawasaki} instead of Metropolis spin flips as used here. This would require some reservoir of up spins to be provided. In the absence of driving, the choice of dynamical update rule has very significant effects on system behaviour \cite{binder_book}: a comparison between Metropolis and Kawasaki dynamics for nucleation under shear would be an interesting topic for further work. Despite its simplicity, the Ising model has made important contributions to the understanding of nucleation phenomena in equilibrium systems, and is likely to yield important insights also for nonequilibrium systems. Moreover, the mechanisms identified here - shear-induced cluster breakup, enhanced cluster growth and cluster coalescence - are likely to play a significant role in experimental nucleation under shear. The discreteness of our model may be a cause for concern: for example, the ``kink'' mechanism for shear-enhanced single spin flip growth illustrated in Figure \ref{fig:kinks}) is a discrete phenomenon. However, our results show that cluster coalescence is of approximately equal importance, and, in any case, preliminary results indicate that the observed qualitative behaviour is not dependent on the size of the largest cluster \cite{avt}. We are encouraged by recent two dimensional Brownian Dynamics simulations of charge stabilised and attractive colloids, in which similar nonmonotonic behaviour of the crystal nucleation rate was observed \cite{cerda}, although the underlying physical mechanisms were postulated to be somewhat different. Future work will investigate the role of transport processes for nucleation under shear in simple model systems and extend our work to more  complex models.

\section*{Acknowledgments}
The authors  thank  G. Bryant, M. Cates, D. Chandler, C. Dellago, E. Vanden-Eijnden, W. Poon, B. Schmittmann and R. Sear for helpful discussions, and K. van Meel for his careful reading of the manuscript. Discussions were stimulated by a workshop at the Erwin Schroedinger Institute (Vienna).  R.J.A. was funded by the Royal Society of Edinburgh, and C.V. was partially funded by EPSRC under grant EP/E030173. This work is part of the research program of the
"Stichting voor Fundamenteel Onderzoek der Materie (FOM)", which is financially supported by the "Nederlandse organisatie voor
Wetenschappelijk Onderzoek (NWO)''. 
\begin{appendix}

\section{The shear algorithm}\label{app:shear}
Here, we describe the algorithm used  to apply shear to the Ising system. The algorithm is similar to that of Cirillo {\em{et al}} \cite{cirillo} but rows are shifted by only one lattice site. The algorithm required the parameters $M_s$ (number of attempted row shifts per row per MC cycle) and $P_s$ (probability of a row shift happening in each attempt) to be assigned, for a given shear rate  $\dot{\gamma}$. This is done as follows: if $\dot{\gamma} \le 1$,  $M_s=1$ and $P_s = \dot{\gamma}$. If $\dot{\gamma} > 1$, then $M_s$ and $P_s$ are allocated such that $M_s$ is the smallest integer value such that $M_s P_s = \dot{\gamma}$ and $0 \le P_s \le 1$. Although this is the strategy used to assign $M_s$ and $P_s$ in this paper, we have verified that our results do not depend on the choice of $M_s$ and $P_s$ for a given $\dot{\gamma}$. 

We then simulate the system using the usual Metropolis Monte Carlo algorithm, except that after each MC cycle, we carry out $M_s \times L$ attempts to shear the system by one lattice site. Each attempt consists of:\\
(1) choose a random number $0 \le r \le 1$\\
(2) If $r \le P_s$, go on to (3)\\
(3) choose a row $j_c$ at random (using a second random number)\\ 
(4) for all spins with y co-ordinate $j > j_c$, carry out the x co-ordinate transformation $i \to i+1$ - taking proper account of periodic boundary conditions\\
(5) update the stored information on y-direction periodic boundary conditions, as detailed below

\begin{figure}[h!]
\begin{center}
{\rotatebox{0}{{\includegraphics[scale=0.4,clip=true]{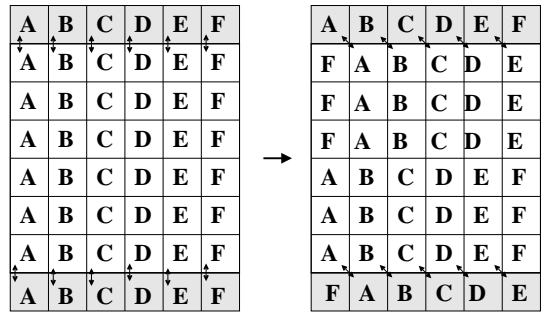}}}}
\caption{{\small Schematic illustration of a row shift in the shearing algorithm. Columns are labelled by letters. The square simulation box is shown flanked by the bottom and top rows of the neighbouring boxes in the up and down direction in the periodic array (shown in grey). The small arrows indicate the neighbours of each lattice site in the top and bottom rows of the box in the up and down directions. A shear step results in a shift of all rows in the top half of the box by one lattice site. This results in a change of identity of the neighbours of the top and bottom rows of the lattice, as shown by the slanting arrows.\label{cirillo_fig}}}
\end{center}
\end{figure}

In the absence of shear, we apply periodic boundary conditions (PBC) in the x and y directions. In the presence of shear, PBC apply in exactly the same way in the x direction, but care is required in the y direction. As illustrated in Figure \ref{cirillo_fig}, a shear step results in a displacement of the identity of the ``up'' neighbours of the top row and the ``down'' neighbours of the bottom row. Therefore, in addition to storing the spin variables $\sigma$ at each lattice site, for every configuration, we also store two integers, $g_{up}$ and $g_{down}$. The identity of the ``up'' neighbour of the spin at position $[i;L]$ (where rows are numbered $1 \to L$) is $[i+g_{up}-nL;1]$, while the identity of the ``down'' neighbour of the spin at position $[i;1]$ is $[i+g_{down}-mL;L]$. The integers $n$ and $m$ are chosen to ensure that $1 \le i+g_{up}-nL \le L$ and  $1 \le i+g_{down}-mL \le L$. Each time a shear step is carried out, $g_{up}$ and $g_{down}$ are updated by $g_{up} \to g_{up}-1$ and  $g_{down} \to g_{down}+1$. It is essential when carrying out the FFS simulations and when reconstructing the transition paths from the FFS sampling that $g_{up}$ and $g_{down}$ are stored for each configuration. 


\section{Sampling the transition state ensemble}\label{app:tse}
To obtain configurations from the transition path ensemble with  $P_B=0.5$ (TSE configurations), we first regenerate the transition paths from the FFS simulation, as described in section \ref{sec:sim}. 
Having regenerated a transition path, we then fire 16 trial runs from every 10th configuration along this path, and monitor the number of trials $N_s^{(1)}$ that reach $B$ before $A$. If $7 \le N_s^{(1)} \le 9$, we then fire a further 100 trial runs from this configuration, and monitor the number trials $N_s^{(2)}$ that reach $B$ rather than $A$. If $40 \le N_s^{(2)} \le 50$, we consider this configuration a member of the $P_B=0.5$ ensemble. This procedure produces a collection of configurations with a range of $P_B$ values around the desired value of 0.5. The parameters of the method can be adjusted to balance computational expense with accuracy. This approach is slightly different from that of Pan and Chandler \cite{pan_jpcb2004}, who include both $N_s^{(1)}$ and $N_s^{(2)}$ in the final computed value of $P_B$.


\end{appendix}


\bibliography{refs_shear}

\end{document}